\documentclass{eptcs}
\usepackage{alltt}
\usepackage{breakurl}

\newcommand{\soft}{SOFT}

\newcommand{\code}[1]{\texttt{#1}}
\newcommand{\icode}[1]{\textit{\code{#1}}}
\newcommand{\codelist}[1]{\code{(}#1\code{)}}
\newcommand{\codecons}[2]{\codelist{#1 \code{.}\ #2}}
\newcommand{\us}{\detokenize{_}}

\newcommand{\funvar}{\icode{fv}}
\newcommand{\funvarI}{\ensuremath{\funvar_1}}
\newcommand{\funvarIP}{\funvarI\code{'}}
\newcommand{\funvarN}[1]{\ensuremath{\funvar_{#1}}}
\newcommand{\funvarNP}[1]{\funvarN{#1}\code{'}}
\newcommand{\var}{\icode{v}}
\newcommand{\varI}{\ensuremath{\var_1}}
\newcommand{\varN}[1]{\ensuremath{\var_{#1}}}
\newcommand{\bvar}{\icode{bv}}
\newcommand{\bvarI}{\ensuremath{\bvar_1}}
\newcommand{\bvarN}[1]{\ensuremath{\bvar_{#1}}}
\newcommand{\sofun}{\icode{sof}}
\newcommand{\sofunP}{\sofun\code{'}}
\newcommand{\sofunPP}{\sofun\code{'}\code{'}}
\newcommand{\sofunWit}{\ensuremath{\sofun_\mathrm{w}}}
\newcommand{\fun}{\icode{f}}
\newcommand{\funP}{\fun\code{'}}
\newcommand{\funPP}{\fun\code{'}\code{'}}
\newcommand{\funI}{\ensuremath{\fun_1}}
\newcommand{\funIP}{\funI\code{'}}
\newcommand{\funN}[1]{\ensuremath{\fun_{#1}}}
\newcommand{\funNP}[1]{\funN{#1}\code{'}}
\newcommand{\funWit}{\ensuremath{\fun_\mathrm{w}}}
\newcommand{\sothm}{\icode{sothm}}
\newcommand{\thm}{\icode{thm}}

\newcommand{\spec}{\icode{spec}}
\newcommand{\specO}{\ensuremath{\spec_0}}
\newcommand{\specI}{\ensuremath{\spec_1}}
\newcommand{\specN}[1]{\ensuremath{\spec_{#1}}}
\newcommand{\funeq}{\code{def}}
\newcommand{\funeqI}{\ensuremath{\funeq_1}}
\newcommand{\funeqN}[1]{\ensuremath{\funeq_{#1}}}

\newcommand{\dotss}{\code{\ldots}}

\newcommand{\funvars}[1]{\mathsf{FV}(#1)}
\newcommand{\inst}{\ensuremath{\Sigma}}
\newcommand{\instof}[1]{\ensuremath{\inst(#1)}}

\newcommand{\secref}[1]{Section~\ref{#1}}

\title{Second-Order Functions and Theorems in ACL2}

\author{Alessandro Coglio
\institute{Kestrel Institute}
\email{\url{http://www.kestrel.edu/~coglio}}
}

\def\titlerunning{Second-Order Functions and Theorems}
\def\authorrunning{A.\ Coglio}
\begin{document}
\maketitle

\begin{abstract}
\soft\ (`Second-Order Functions and Theorems') is a tool
to mimic second-order functions and theorems in the first-order logic of ACL2.
Second-order functions are mimicked
by first-order functions that reference
explicitly designated uninterpreted functions that mimic function variables.
First-order theorems over these second-order functions
mimic second-order theorems universally quantified over function variables.
Instances of second-order functions and theorems
are systematically generated
by replacing function variables with functions.
\soft\ can be used to carry out program refinement inside ACL2,
by constructing a sequence of increasingly stronger second-order predicates
over one or more target functions:
the sequence starts with a predicate
that specifies requirements for the target functions,
and ends with a predicate that
provides executable definitions for the target functions.
\end{abstract}

\section{The \soft\ Tool}
\label{sec:soft}

\soft\ (`Second-Order Functions and Theorems')
is a tool to mimic second-order functions and theorems~\cite{andrews-logic}
in the first-order logic of ACL2~\cite{acl2-www}.
Second-order functions are mimicked
by first-order functions that reference
explicitly designated uninterpreted functions that mimic function variables.
First-order theorems over these second-order functions
mimic second-order theorems universally quantified over function variables.
Instances of second-order functions and theorems
are systematically generated
by replacing function variables with functions.
Theorem instances are proved automatically,
via automatically generated
functional instantiations~\cite{boyer-instantiation}.

\soft\ does not extend the ACL2 logic.
It is an ACL2 library,
available in the ACL2 community books,
that provides macros to introduce
function variables,
second-order functions,
second-order theorems,
and instances thereof.
The macros modify the ACL2 state
only by submitting sound and conservative events;
they cannot introduce unsoundness or inconsistency on their own.
The main features of the macros are described and exemplified below;
full details are in their documentation and implementation.

\subsection{Function Variables}
\label{sec:funvars}

A \emph{function variable} is introduced as
\begin{alltt}
  (defunvar \funvar\ (* \ldots\ *) => *)
\end{alltt}
where:
\begin{itemize}
\item
  \funvar\ is a symbol, which names the function variable.
\item
  \codelist{\code{*} \dotss\ \code{*}} is a list of 1 or more \code{*}s,
  which defines the arity, i.e.\ type~\cite{church-types}, of \funvar.
\end{itemize}
This generates the event
\begin{alltt}
  (defstub \funvar\ (* \ldots\ *) => *)
\end{alltt}
i.e.\ \funvar\ is introduced as an uninterpreted function with the given type.
Furthermore, a \code{table} event is generated
to record \funvar\ in a global table of function variables.

For example,
\begin{alltt}
  (defunvar ?f (*) => *)
  (defunvar ?p (*) => *)
  (defunvar ?g (* *) => *)
\end{alltt}
introduce two unary function variables and one binary function variable.
Starting function variable names with \code{?}
provides a visual cue for their function variable status,
but \soft\ does not enforce this naming convention.

\subsection{Second-Order Functions}
\label{sec:sofuns}

\soft\ supports three kinds of second-order functions:
plain second-order functions,
choice second-order functions,
and quantifier second-order functions.

\subsubsection{Plain Functions}
\label{sec:plain}

A \emph{plain second-order function} is introduced as
\begin{alltt}
  (defun2 \sofun\ (\funvarI\ \ldots\ \funvarN{n}) (\varI\ \ldots \varN{m}) \icode{doc} \icode{decl} \ldots \icode{decl} \icode{body})
\end{alltt}
where:
\begin{itemize}
\item
  \sofun\ is a symbol, which names the second-order function.
\item
  \codelist{\funvarI\ \dotss\ \funvarN{n}}
  is a non-empty list without duplicates
  of previously introduced function variables,
  whose order is immaterial,
  which are the function parameters of \sofun.
\item
  The other items are as in \code{defun}:
  individual variables,
  optional documentation string,
  optional declarations,
  and defining body.
\item
  $\funvars{\icode{body}}
   \ \cup\
   \funvars{\icode{measure}}
   \ \cup\
   \funvars{\icode{guard}}
   \ =\
   \{\funvarI, \ldots, \funvarN{n}\}$,
  where:
  \begin{itemize}
  \item
    \icode{measure} is the measure expression of \sofun,
    or \code{nil} if \sofun\ is not recursive.
  \item
    \icode{guard} is the guard of \sofun\
    (\code{t} if not given explicitly in the declarations).
  \item
    $\funvars{\icode{term}}$ is the set of function variables
    that either occur in \icode{term}
    or are function parameters of second-order functions
    that occur in \icode{term}.
  \end{itemize}
  I.e.\ the function parameters of \sofun\
  are all and only the function variables that \sofun\ depends on.%
  \footnote{Thus, \code{defun2} could have been defined
  to have the same form as \code{defun},
  i.e.\ without \codelist{\funvarI\ \dotss\ \funvarN{n}}.
  However, the presence of the functions parameters parallels
  that of the individual parameters,
  and the redundancy check may detect user errors.}
\end{itemize}
This generates the event
\begin{alltt}
  (defun \sofun\ (\varI\ \ldots \varN{m}) \icode{doc} \icode{decl} \ldots \icode{decl} \icode{body})
\end{alltt}
i.e.\ \sofun\ is introduced as a first-order function using \code{defun},
removing the function variables.
Furthermore, a \code{table} event is generated
to record \sofun\ in a global table of second-order functions.

For example,
\begin{alltt}
  (defun2 quad[?f] (?f) (x)
    (?f (?f (?f (?f x)))))
\end{alltt}
introduces a non-recursive function
to apply its function parameter to its individual parameter four times.
The name \code{quad[?f]}
conveys the dependency on the function parameter
and provides a visual cue for the implicit presence of the function parameter
when the function is applied, e.g.\ in \codelist{\code{quad[?f]} \code{x}},
but \soft\ does not enforce this naming convention.

As another example,
\begin{alltt}
  (defun2 all[?p] (?p) (l)
    (cond ((atom l) (null l))
          (t (and (?p (car l)) (all[?p] (cdr l))))))
\end{alltt}
introduces a recursive predicate (i.e.\ boolean-valued function)
that recognizes \code{nil}-terminated lists
whose elements satisfy the predicate parameter.

As a third example,
\begin{alltt}
  (defun2 map[?f_?p] (?f ?p) (l)
    (declare (xargs :guard (all[?p] l)))
    (cond ((endp l) nil)
          (t (cons (?f (car l)) (map[?f_?p] (cdr l))))))
\end{alltt}
introduces a recursive function that homomorphically lifts \code{?f}
to operate on \code{nil}-terminated lists whose elements satisfy \code{?p}.
The predicate parameter \code{?p} only appears in the guard, not in the body.

As a fourth example,
\begin{alltt}
  (defun2 fold[?f_?g] (?f ?g) (bt)
    (cond ((atom bt) (?f bt))
          (t (?g (fold[?f_?g] (car bt)) (fold[?f_?g] (cdr bt))))))
\end{alltt}
introduces a generic folding function on values as binary trees.

\subsubsection{Choice Functions}
\label{sec:choice}

A \emph{choice second-order function} is introduced as
\begin{alltt}
  (defchoose2 \sofun\ (\bvarI \ldots\ \bvarN{p}) (\funvarI\ \ldots\ \funvarN{n}) (\varI\ \ldots \varN{m}) \icode{body} \icode{key-opts})
\end{alltt}
where:
\begin{itemize}
\item
  \sofun\ is a symbol, which names the second-order function.
\item
  \codelist{\funvarI\ \dotss\ \funvarN{n}} are the function parameters,
  as in \code{defun2}.
\item
  The other items are as in \code{defchoose}:
  bound variables,
  individual variables,
  constraining body,
  and keyed options.
\item
  $\funvars{\icode{body}}\ =\ \{\funvarI, \ldots, \funvarN{n}\}$.
\end{itemize}
This generates the event
\begin{alltt}
  (defchoose \sofun\ (\bvarI \ldots\ \bvarN{p}) (\varI\ \ldots \varN{m}) \icode{body} \icode{key-opts})
\end{alltt}
i.e.\ \sofun\ is introduced as a first-order function using \code{defchoose},
removing the function variables.
Furthermore, a \code{table} event is generated to record \sofun\
in the same global table where plain second-order functions are recorded.

For example,
\begin{alltt}
  (defchoose2 fixpoint[?f] x (?f) ()
    (equal (?f x) x))
\end{alltt}
introduces a second-order function
constrained to return a fixed point of \code{?f}, if any exists.

\subsubsection{Quantifier Functions}
\label{sec:quant}

A \emph{quantifier second-order function} is introduced as
\begin{alltt}
  (defun-sk2 \sofun\ (\funvarI\ \ldots\ \funvarN{n}) (\varI\ \ldots \varN{m}) \icode{body} \icode{key-opts})
\end{alltt}
where:
\begin{itemize}
\item
  \sofun\ is a symbol, which names the second-order function.
\item
  \codelist{\funvarI\ \dotss\ \funvarN{n}} are the function parameters,
  as in \code{defun2} and \code{defchoose2}.
\item
  The other items are as in \code{defun-sk}:
  individual variables,
  defining body,
  and keyed options.
\item
  $\funvars{\icode{body}}
   \ \cup\
   \funvars{\icode{guard}}
   \ =\
   \{\funvarI, \ldots, \funvarN{n}\}$,
   where \icode{guard} is the guard of \sofun\
   (\code{t} if not given explicitly in the \code{:witness-dcls} option).
\end{itemize}
This generates the event
\begin{alltt}
  (defun-sk \sofun\ (\varI\ \ldots \varN{m}) \icode{body} \icode{key-opts})
\end{alltt}
i.e.\ \sofun\ is introduced as a first-order function using \code{defun-sk},
removing the function variables.
Furthermore, a \code{table} event is generated
to record \sofun\ in the same global table
where plain and choice second-order functions are recorded.

For example,
\begin{alltt}
  (defun-sk2 injective[?f] (?f) ()
    (forall (x y) (implies (equal (?f x) (?f y)) (equal x y))))
\end{alltt}
introduces a predicate that recognizes injective functions.

\subsection{Instances of Second-Order Functions}
\label{sec:sofuninst}

An \emph{instance of a second-order function} is a function introduced as
\begin{alltt}
  (defun-inst \fun\ (\funvarI\ \ldots\ \funvarN{n}) (\sofun\ (\funvarI' . \funI') \ldots\ (\funvarN{m}' . \funN{m}')) \icode{key-opts})
\end{alltt}
where:
\begin{itemize}
\item
  \fun\ is a symbol, which names the new function.
\item
  \codelist{\funvarI\ \dotss\ \funvarN{n}} are optional function parameters.
  If present, \fun\ is a second-order function;
  if absent, \fun\ is a first-order function.
\item
  \sofun\ is a previously introduced second-order function.
\item
  \codelist{\codecons{\funvarIP}{\funIP}
            \dotss\
            \codecons{\funvarNP{m}}{\funNP{m}}}
  is an \emph{instantiation} \inst,
  i.e.\ an alist whose keys \funvarNP{i} are distinct function variables,
  whose values \funNP{i} are previously introduced
  function variables, second-order functions, or regular first-order functions,
  and where each \funNP{i} has the same type as \funvarNP{i}.
  Each \funvarNP{i} is a function parameter of \sofun.
  The notation
  \codelist{\sofun\
            \codecons{\funvarIP}{\funIP}
            \dotss\
            \codecons{\funvarNP{m}}{\funNP{m}}}
  suggests the application of \sofun\ to the functions \funNP{i};
  since the function parameters of \sofun\ are unordered,
  the application is by explicit association, not positional.
  An instance of a second-order function is introduced
  as a named application of the second-order function;
  \soft\ does not support the application of a second-order function
  on the fly within a term, as in the application of a first-order function.
  Not all the function parameters of \sofun\ must be keys in \inst;
  missing function parameters are left unchanged.
\item
  \icode{key-opts} are keyed options,
  e.g.\ to override attributes of \fun\ that are otherwise derived from \sofun.
\item
  If \sofun\ is a plain function,
  $\funvars{\instof{\icode{body}}}
   \ \cup\
   \funvars{\instof{\icode{measure}}}
   \ \cup\
   \funvars{\instof{\icode{guard}}}
   \ =\
   \{\funvarI, \ldots, \funvarN{n}\}$,
  where \icode{body}, \icode{measure}, and \icode{guard}
  are the body,
  measure expression (\code{nil} if \sofun\ is not recursive),
  and guard
  of \sofun,
  and \instof{\icode{term}} is the result of applying \inst\ to \icode{term}
  (see below).
\item
  If \sofun\ is a choice function,
  $\funvars{\instof{\icode{body}}} = \{\funvarI, \ldots, \funvarN{n}\}$,
  where \icode{body} is the body of \sofun.
\item
  If \sofun\ is a quantifier function,
  $\funvars{\instof{\icode{body}}}
   \ \cup\
   \funvars{\instof{\icode{guard}}}
   \ =\
   \{\funvarI, \ldots, \funvarN{n}\}$,
  where \icode{body} and \icode{guard} are the body and guard of \sofun.
\end{itemize}
This generates a \code{defun}, \code{defchoose}, or \code{defun-sk} event,
depending on whether
\sofun\ is a plain, choice, or quantifier function.
The event introduces \fun\
with body \instof{\icode{body}},
measure \instof{\icode{measure}} (if \sofun\ is recursive, hence plain),
and guard \instof{\icode{guard}} (if \sofun\ is a plain or quantifier function).
\fun\ is recursive iff\ \sofun\ is recursive:
\code{defun-inst} generates the termination proof of \fun\
from the termination proof of \sofun\
using the techniques to instantiate second-order theorems
described in \secref{sec:sothminst}.

Furthermore, \code{defun-inst} generates a \code{table} event
to record \fun\ as the \inst\ instance of \sofun\
in a global table of instances of second-order functions.
If \fun\ is second-order, \code{defun-inst} also generates a \code{table} event
to record \fun\ in the global table of second-order functions.

\instof{\icode{term}} is obtained from \icode{term}
by replacing the keys of \inst\ in \icode{term} with their values in \inst.
This involves not only explicit occurrences of such keys in \icode{term},
but also implicit occurrences as function parameters
of second-order functions occurring in \icode{term}.
For example,
if the pair \codecons{\code{?f}}{\code{f}} is in \inst,
\code{sof[...?f...]} is a second-order function
whose function parameters include \code{?f},
and \icode{term} is
\codelist{\code{cons} \codelist{\code{?f} \code{x}}
                      \codelist{\code{sof[...?f...]} \code{y}}},
then \instof{\icode{term}} is
\codelist{\code{cons} \codelist{\code{f} \code{x}}
                      \codelist{\code{sof[...f...]} \code{y}}},
where \code{sof[...f...]} is the \inst' instance of \code{sof[...?f...]},
where \inst' is the restriction of \inst\
to the keys that are function parameters of \code{sof[...?f...]}.
The table of instances of second-order functions
is consulted to find \code{sof[...f...]}.
If the instance is not in the table,
\code{defun-inst} fails:
the user must introduce \code{sof[...f...]}, via a \code{defun-inst},
and then re-try the failed instantiation.

For example, given a function
\begin{alltt}
  (defun wrap (x) (list x))
\end{alltt}
that wraps a value into a singleton list,
\begin{alltt}
  (defun-inst quad[wrap]
    (quad[?f] (?f . wrap)))
\end{alltt}
introduces a function that wraps a value four times.

As another example, given a predicate
\begin{alltt}
  (defun octetp (x) (and (natp x) (< x 256)))
\end{alltt}
that recognizes octets,
\begin{alltt}
  (defun-inst all[octetp]
    (all[?p] (?p . octetp)))
\end{alltt}
introduces a predicate that recognizes \code{nil}-terminated lists of octets.

As a third example,
\begin{alltt}
  (defun-inst map[code-char]
    (map[?f_?p] (?f . code-char) (?p . octetp)))
\end{alltt}
introduces a function
that translates lists of octets to lists of corresponding characters.
The replacement \code{code-char} of \code{?f}
induces the replacement \code{octetp} of \code{?p},
because the guard of \code{code-char} is (equivalent to) \code{octetp};
the name \code{map[code-char]} indicates
only the replacement of \code{?f} explicitly.

As a fourth example,
\begin{alltt}
  (defun-inst fold[nfix_plus]
    (fold[?f_?g] (?f . nfix) (?g . binary-+)))
\end{alltt}
adds up all the natural numbers in a tree, coercing other values to 0.

As a fifth example, given a function
\begin{alltt}
  (defun twice (x) (* 2 (fix x)))
\end{alltt}
that doubles a value,
\begin{alltt}
  (defun-inst fixpoint[twice]
    (fixpoint[?f] (?f . twice)))
\end{alltt}
introduces a function constrained to return
the (only) fixed point 0 of \code{twice}.

As a sixth example,
\begin{alltt}
  (defun-inst injective[quad[?f]] (?f)
    (injective[?f] (?f . quad[?f])))
\end{alltt}
introduces a predicate that recognizes functions
whose four-fold application is injective.

\subsection{Second-Order Theorems}
\label{sec:sothms}

A \emph{second-order theorem} is a theorem
whose formula depends on function variables,
which occur in the theorem
or are function parameters of second-order functions that occur in the theorem.
Since function variables are unconstrained,
a second-order theorem is effectively
universally quantified over the function variables that it depends on.
It is introduced via standard events like \code{defthm}.%
\footnote{The absence of an explicit quantification over function variables
in second-order theorems
parallels the absence of an explicit quantification over individual variables
in first-order theorems.}

For example,
\begin{alltt}
  (defthm len-of-map[?f_?p]
    (equal (len (map[?f_?p] l)) (len l)))
\end{alltt}
shows that the homomorphic lifting of \code{?f} to lists of \code{?p} values
preserves the length of the list,
for every function \code{?f} and predicate \code{?p}.

As another example,
\begin{alltt}
  (defthm injective[quad[?f]]-when-injective[?f]
    (implies (injective[?f]) (injective[quad[?f]]))
    :hints
    (("Goal" :use
      ((:instance
        injective[?f]-necc
        (x (?f (?f (?f (?f (mv-nth 0 (injective[quad[?f]]-witness)))))))
        (y (?f (?f (?f (?f (mv-nth 1 (injective[quad[?f]]-witness))))))))
       (:instance
        injective[?f]-necc
        (x (?f (?f (?f (mv-nth 0 (injective[quad[?f]]-witness))))))
        (y (?f (?f (?f (mv-nth 1 (injective[quad[?f]]-witness)))))))
       (:instance
        injective[?f]-necc
        (x (?f (?f (mv-nth 0 (injective[quad[?f]]-witness)))))
        (y (?f (?f (mv-nth 1 (injective[quad[?f]]-witness))))))
       (:instance
        injective[?f]-necc
        (x (?f (mv-nth 0 (injective[quad[?f]]-witness))))
        (y (?f (mv-nth 1 (injective[quad[?f]]-witness)))))
       (:instance
        injective[?f]-necc
        (x (mv-nth 0 (injective[quad[?f]]-witness)))
        (y (mv-nth 1 (injective[quad[?f]]-witness))))))))
\end{alltt}
shows that the four-fold application of an injective function is injective.

As a third example,
given a function variable
\begin{alltt}
  (defunvar ?io (* *) => *)
\end{alltt}
for an abstract input/output relation,
a predicate
\begin{alltt}
  (defun-sk2 atom-io[?f_?io] (?f ?io) ()
    (forall x (implies (atom x) (?io x (?f x))))
    :rewrite :direct)
\end{alltt}
that recognizes functions \code{?f}
that satisfy the input/output relation on atoms,
and a predicate
\begin{alltt}
  (defun-sk2 consp-io[?g_?io] (?g ?io) ()
    (forall (x y1 y2)
            (implies (and (consp x) (?io (car x) y1) (?io (cdr x) y2))
                     (?io x (?g y1 y2))))
    :rewrite :direct)
\end{alltt}
that recognizes functions \code{?g}
that satisfy the input/output relation on \code{cons} pairs
when the arguments are valid outputs
for the \code{car} and \code{cdr} components,
\begin{alltt}
  (defthm fold-io[?f_?g_?io]
    (implies (and (atom-io[?f_?io]) (consp-io[?g_?io]))
             (?io x (fold[?f_?g] x))))
\end{alltt}
shows that the generic folding function on binary trees
satisfies the input/output relation
when its function parameters satisfy the predicates just introduced.

\subsection{Instances of Second-Order Theorems}
\label{sec:sothminst}

An \emph{instance of a second-order theorem} is a theorem introduced as
\begin{alltt}
  (defthm-inst \thm\ (\sothm (\funvarI . \funI) \ldots\ (\funvarN{n} . \funN{n})) :rule-classes \ldots)
\end{alltt}
where:
\begin{itemize}
\item
  \thm\ is a symbol, which names the new theorem.
\item
  \sothm\ is a previously introduced second-order theorem.
\item
  \codelist{\codecons{\funvarI}{\funI}
            \dotss\
            \codecons{\funvarN{n}}{\funN{n}}}
  is an instantiation \inst,
  where each \funvarN{i} is a function variable that \sothm\ depends on.
  The notation
  \codelist{\sothm\
            \codecons{\funvarI}{\funI}
            \dotss\
            \codecons{\funvarN{m}}{\funN{m}}}
  is similar to \code{defun-inst}.
\item
  The keyed option \code{:rule-classes} \dotss\ is as in \code{defthm}.
\end{itemize}
This generates the event
\begin{alltt}
  (defthm \thm\ \instof{\icode{formula}} :rule-classes ... :instructions \icode{proof})
\end{alltt}
where:
\begin{itemize}
\item
  \icode{formula} is the formula of \sothm.
\item
  \icode{proof} consists of two commands for the ACL2 proof checker
  to prove \thm\ using \sothm.
\end{itemize}

The first command of \icode{proof} is
\begin{alltt}
  (:use (:functional-instance \sothm (\funvarI\ \funI) ... (\funvarN{n} \funN{n}) \icode{more-pairs}))
\end{alltt}
i.e.\ \thm\ is proved using a functional instance of \sothm.
The pairs that define the functional instance include
not only the pairs that form \inst\
(in list notation instead of dotted notation),
but also, in \icode{more-pairs} above,
all the pairs \codelist{\sofun\ \fun} such that
\sofun\ is a second-order function that occurs in \sothm\
and \fun\ is its replacement in \thm\
(i.e.\ \fun\ is the \inst' instance of \sofun,
where \inst' is the restriction of \inst\ to the function parameters of \sofun).
These additional pairs are determined in the same way
as when \inst\ is applied to \icode{formula}
(see \secref{sec:sofuninst}):
thus, the result of \codelist{\code{:functional-instance} \dotss} above
is \instof{\icode{formula}},
and the main goal of \thm\ is readily proved.

The use of the functional instance reduces the proof of \thm\
to proving that, for each pair,
the replacing function satisfies all the constraints of the replaced function.
Since function variables are unconstrained,
nothing needs to be proved for the \codelist{\funvarN{i} \funN{i}} pairs.
For each \codelist{\sofun\ \fun} pair in \icode{more-pairs},
it must be proved that \fun\ satisfies the constraints on \sofun.
If \sofun\ references another second-order function \sofunP\
that depends on some \funvarN{i},
a further pair \codelist{\sofunP\ \funP} goes into \icode{more-pairs},
where \funP\ is the appropriate instance of \sofunP,
so that the constraints on \sofun\ to be proved are properly instantiated.
This further pair generates further constraints to be proved.
To properly instantiate these further constraints,
another pair \codelist{\sofunPP\ \funPP} goes into \icode{more-pairs},
if \sofunPP\ is a second-order function referenced by \sofunP\
that depends on some \funvarN{i},
and \funPP\ is the appropriate instance of \sofunPP.
Therefore, \icode{more-pairs} includes all the pairs \codelist{\sofun\ \fun}
such that \sofun\ is a second-order function
that is directly or indirectly referenced by \sothm\
and that depends on some \funvarN{i},
and \fun\ is the appropriate instance of \sofun.

If \sofun\ is a quantifier second-order function,
it references a witness function \sofunWit\ introduced by \code{defun-sk}.
The \code{defun-sk} that introduces the instance \fun\ of \sofun\
also introduces a witness function \funWit\
that is effectively an instance of \sofunWit,
but is not recorded in the table of instances of second-order functions
because \sofunWit\ and \funWit\ are ``internal''.
The pair \codelist{\sofunWit\ \funWit} goes into \icode{more-pairs} as well.

For each pair \codelist{\sofun\ \fun} in \icode{more-pairs},
the constraints of \sofun\ are:
the definition of \sofun\ if \sofun\ is a plain function;
the constraining axiom of \sofun\ if \sofun\ is a choice function;
the definition of \sofun\ and the rewrite rule of \sofun\
if \sofun\ is a quantifier function
(the rewrite rule of \sofun\ is generated by \code{defun-sk};
its default name is
\sofun\code{-necc} if the quantifier is universal,
\sofun\code{-suff} if the quantifier is existential).
Instantiating these constraints yields
the corresponding definitions, constraining axioms, and rewrite rules of \fun,
by the construction of the instance \fun\ of \sofun.

The second command of \icode{proof} is
\begin{alltt}
  (:repeat (:then (:use \icode{facts}) :prove))
\end{alltt}
where \icode{facts} includes
the names of all the \fun\ functions in \icode{more-pairs},
which are also the names of their definitions and constraining axioms;
\icode{facts} also includes
the names of the rewrite rules for quantifier functions.
This command runs the prover on every proof subgoal,
after augmenting each subgoal with all the facts in \icode{facts}.
This command has worked on all the examples tried so far,
but a more honed approach could be investigated,
should some future example fail;
since the constraints are satisfied by construction,
this is just an implementation issue.

For example,
\begin{alltt}
  (defthm-inst len-of-map[code-char]
    (len-of-map[?f_?p] (?f . code-char) (?p . octetp)))
\end{alltt}
shows that \code{map[code-char]} preserves the length of the list.

As another example,
given instances
\begin{alltt}
  (defun-inst injective[quad[wrap]] (injective[quad[?f]] (?f . wrap)))
  (defun-inst injective[wrap] (injective[?f] (?f . wrap)))
\end{alltt}
the theorem instance
\begin{alltt}
  (defthm-inst injective[quad[wrap]]-when-injective[wrap]
    (injective[quad[?f]]-when-injective[?f] (?f . wrap)))
\end{alltt}
shows that \code{quad[wrap]} is injective if \code{wrap} is.

An example instance of \code{fold-io[?f\us?g\us?io]}
is in \secref{sec:progref}.

\subsection{Summary of the Macros}
\label{sec:summary}

\code{defunvar}, \code{defun2}, \code{defchoose2}, and \code{defun-sk2}
are wrappers of existing events
that explicate function variable dependencies
and record additional information.
They set the stage for \code{defun-inst} and \code{defthm-inst}.

\code{defun-inst} provides the ability to concisely generate functions,
and automatically prove their termination if recursive,
by specifying replacements of function variables.

\code{defthm-inst} provides the ability to
concisely generate and automatically prove theorems,
by specifying replacements of function variables.

\section{Use in Program Refinement}
\label{sec:progref}

In program refinement~\cite{dijkstra-constructive},
a correct-by-construction implementation
is derived from a requirements specification
via a sequence of intermediate specifications.
\emph{Shallow pop-refinement}
(where `pop' stands for `predicates over programs')
is an approach to program refinement,
carried out inside an interactive theorem prover
by constructing a sequence of increasingly stronger predicates
over one or more target functions.
The sequence starts with a predicate
that specifies requirements for the target functions,
and ends with a predicate that
provides executable definitions for the target functions.
Shallow pop-refinement is a form of pop-refinement~\cite{coglio-pop}
in which the programs predicated upon are
shallowly embedded functions of the logic of the theorem prover,
instead of deeply embedded programs of a programming language
as in~\cite{coglio-pop}.

\soft\ can be used to carry out shallow pop-refinement in ACL2,
as explained and exemplified below.
The example derivation is overkill for the simple program obtained,
which can be easily written and proved correct directly.
But the purpose of the example is to illustrate
techniques that can be used to derive more complex programs,
and how \soft\ supports these techniques
(which are more directly supported in higher-order logic).
The hints in some of the theorems below distill their proofs
into patterns that should apply to similarly structured derivations,
suggesting opportunities for future automation.

\subsection{Specifications as Second-Order Predicates}
\label{sec:progref-spec}

Requirements over $n \geq 1$ target functions are specified
by introducing function variables $\funvarI,\ldots,\funvarN{n}$
that represent the target functions,
and by defining a second-order predicate \specO\
over $\funvarI,\ldots,\funvarN{n}$
that asserts the required properties of the target functions.
The possible implementations
are all the $n$-tuples of executable functions that satisfy the predicate.
The task is to find such an $n$-tuple,
thus constructively proving the predicate,
existentially quantified over the function parameters.

For example,
given a function
\begin{alltt}
  (defun leaf (e bt)
    (cond ((atom bt) (equal e bt))
          (t (or (leaf e (car bt)) (leaf e (cdr bt))))))
\end{alltt}
to test whether something is a leaf of a binary tree,
a function to extract from a binary tree the leaves that are natural numbers,
in no particular order and possibly with duplicates,
can be specified as
\begin{alltt}
  (defunvar ?h (*) => *)
  (defun-sk io (x y) ; input/output relation
    (forall e (iff (member e y) (and (leaf e x) (natp e))))
    :rewrite :direct)
  (defun-sk2 spec[?h] (?h) ()
    (forall x (io x (?h x)))
    :rewrite :direct)
\end{alltt}
The task is to solve \code{spec[?h]} for \code{?h},
i.e.\ to find an executable function \code{h}
such that the instance \code{spec[h]} of \code{spec[?h]} holds.

Properties implied by the requirements are proved
as second-order theorems with \specO\ as hypothesis,
e.g.\ for validation purposes.
Since the function parameters are universally quantified in the theorem,
the properties hold for all the implementations of the specification.

For example,
the members of the output of every implementation of \code{spec[?h]}
are natural numbers:
\begin{alltt}
  (defthm natp-of-member-of-output
    (implies (and (spec[?h]) (member e (?h x))) (natp e))
    :hints (("Goal" :use (spec[?h]-necc (:instance io-necc (y (?h x)))))))
\end{alltt}

\subsection{Refinement as Second-Order Predicate Strengthening}
\label{sec:progref-ref}

The specification \specO\ is stepwise refined
by constructing a sequence $\specI,\ldots,\specN{m}$
of increasingly stronger predicates over $\funvarI,\ldots,\funvarN{n}$.
Each such predicate embodies a decision that
either narrows down the possible implementations
or rephrases their description towards their determination.
The correctness of each step $j\in\{1,\ldots,m\}$
is expressed by the second-order theorem
\codelist{\code{implies} \codelist{\specN{j}} \codelist{\specN{j-1}}}.

The sequence ends with $\specN{m}$ asserting that
each \funvarN{i} is equal to some executable function \funN{i}:%
\footnote{The body of each \codelist{\code{defun-sk2} \funeqN{i} \dotss}
is a first-order expression
of the second-order equality $\funvarN{i} = \funN{i}$.}
\begin{alltt}
  (defun-sk2 \funeqI\ (\funvarI) () (forall x (equal (\funvarI x) (\funI x))))
  ...
  (defun-sk2 \funeqN{n}\ (\funvarN{n}) () (forall x (equal (\funvarN{n} x) (\funN{n} x))))
  (defun2 \specN{m} (\funvarI\ ... \funvarN{n}) () (and (\funeqI) ... (\funeqN{n})))
\end{alltt}
The tuple $\langle\funI,\ldots,\funN{n}\rangle$ is the implementation.
Chaining the implications of the $m$ step correctness theorems
yields the second-order theorem
\codelist{\code{implies} \codelist{\specN{m}} \codelist{\specO}}.
Its \inst\ instance, where \inst\ is the instantiation
\codelist{\codecons{\funvarI}{\funI} \dotss\ \codecons{\funvarN{n}}{\funN{n}}},
is essentially \instof{\codelist{\specO}}
(because \instof{\codelist{\specN{m}}} is trivially true),
which asserts that
the implementation $\langle\funI,\ldots,\funN{n}\rangle$ satisfies \specO.

More precisely,
in the course of the derivation,
function variables $\funvarN{n+1},\ldots,\funvarN{n+p}$ may be added
to represent additional target functions $\funN{n+1},\ldots,\funN{n+p}$
called by $\funI,\ldots,\funN{n}$.
This may happen as the task of finding $\funI,\ldots,\funN{n}$
is progressively reduced to simpler sub-tasks
of finding $\funN{n+1},\ldots,\funN{n+p}$.
If $\funvarN{n+k}$ is added at refinement step $j$,
since $\specN{j-1}$ does not depend on $\funvarN{n+k}$,
the universal quantification of $\funvarN{n+k}$
over the step correctness theorem
\codelist{\code{implies} \codelist{\specN{j}} \codelist{\specN{j-1}}}
is equivalent to an existential quantification of $\funvarN{n+k}$
over the hypothesis \codelist{\specN{j}} of the theorem.
The complete implementation that results from the derivation is
$\langle\funI,\ldots,\funN{n},\funN{n+1},\ldots,\funN{n+p}\rangle$.

The function variables \funvarN{i}
are placeholders for the target functions in the \specN{j} predicates.
Each \funvarN{i} remains uninterpreted throughout the derivation;
no constraints are attached to it via axioms.
Each \specN{j} is defined, so it does not introduce logical inconsistency.
Inconsistent requirements on the target functions
amount to \specO\ being always false, not to logical inconsistency.
Obtaining an implementation witnesses the consistency of the requirements.

For example,
\code{spec[?h]} from \secref{sec:progref-spec} can be refined as follows.

\paragraph{Step 1}

Since the target function represented by \code{?h} operates on binary trees,
\code{spec[?h]} is strengthened by constraining \code{?h}
to be the folding function on binary trees from \secref{sec:plain}:
\begin{alltt}
  (defun-sk2 def-?h-fold[?f_?g] (?h ?f ?g) ()
    (forall x (equal (?h x) (fold[?f_?g] x)))
    :rewrite :direct)
  (defun2 spec1[?h_?f_?g] (?h ?f ?g) ()
    (and (def-?h-fold[?f_?g]) (spec[?h])))
  (defthm step1 (implies (spec1[?h_?f_?g]) (spec[?h]))
    :hints (("Goal" :in-theory '(spec1[?h_?f_?g]))))
\end{alltt}
The predicate \code{spec1[?h\us?f\us?g]} adds to \code{spec[?h]}
the conjunct \code{def-?h-fold[?f\us?g]}.
Thus, the task of finding a solution for \code{?h} is reduced
to the task of finding solutions for \code{?f} and \code{?g}:
instantiating \code{def-?h-fold[?f\us?g]}
with solutions for \code{?f} and \code{?g} yields a solution for \code{?h},
in Step 5 below.

\paragraph{Step 2}

The theorem \code{fold-io[?f\us?g\us?io]} from \secref{sec:sothms},
which shows the correctness of the folding function
(with respect to an input/output relation)
under suitable correctness assumptions on the function parameters,
is instantiated with the input/output relation \code{io}
used in \code{spec[?h]}:
\begin{alltt}
  (defun-inst atom-io[?f] (?f) (atom-io[?f_?io] (?io . io)))
  (defun-inst consp-io[?g] (?g) (consp-io[?g_?io] (?io . io)))
  (defthm-inst fold-io[?f_?g] (fold-io[?f_?g_?io] (?io . io)))
\end{alltt}
Since the conclusion \code{(io x (fold[?f\us?g] x))} of \code{fold-io[?f\us?g]}
equals the matrix \code{(io x (?h x))} of \code{spec[?h]}
when \code{def-?h-fold[?f\us?g]} holds,
\code{spec1[?h\us?f\us?g]} is strengthened
by replacing the \code{spec[?h]} conjunct
with the hypotheses of \code{fold-io[?f\us?g]}:
\begin{alltt}
  (defun2 spec2[?h_?f_?g] (?h ?f ?g) ()
    (and (def-?h-fold[?f_?g]) (atom-io[?f]) (consp-io[?g])))
  (defthm step2 (implies (spec2[?h_?f_?g]) (spec1[?h_?f_?g]))
    :hints (("Goal" :in-theory '(spec1[?h_?f_?g] spec2[?h_?f_?g] spec[?h]
                                 def-?h-fold[?f_?g]-necc fold-io[?f_?g]))))
\end{alltt}

\paragraph{Step 3}

The predicate \code{atom-io[?f]} specifies requirements on \code{?f}
independently from \code{?g} and \code{?h}.
An implementation \code{f} can be derived
by constructing a sequence of increasingly stronger predicates over \code{?f},
in the same way in which \code{spec[?h]} is being refined stepwise.
This is a possible final result:
\begin{alltt}
  (defun f (x) (if (natp x) (list x) nil))
  (defun-inst atom-io[f] (atom-io[?f] (?f . f)))
  (defthm atom-io[f]! (atom-io[f]))
\end{alltt}
The predicate \code{spec2[?h\us?f\us?g]} is strengthened
by replacing the \code{atom-io[?f]} conjunct
with one that constrains \code{?f} to be \code{f}:
\begin{alltt}
  (defun-sk2 def-?f (?f) () (forall x (equal (?f x) (f x))) :rewrite :direct)
  (defun2 spec3[?h_?f_?g] (?h ?f ?g) ()
    (and (def-?h-fold[?f_?g]) (def-?f) (consp-io[?g])))
  (defthm step3-lemma (implies (def-?f) (atom-io[?f]))
    :hints (("Goal" :in-theory '(atom-io[?f] atom-io[f]-necc
                                 atom-io[f]! def-?f-necc))))
  (defthm step3 (implies (spec3[?h_?f_?g]) (spec2[?h_?f_?g]))
    :hints (("Goal" :in-theory '(spec2[?h_?f_?g] spec3[?h_?f_?g] step3-lemma))))
\end{alltt}

\paragraph{Step 4}

The predicate \code{consp-io[?g]} specifies requirements on \code{?g}
independently from \code{?f} and \code{?h}.
An implementation \code{g} can be derived
by constructing a sequence of increasingly stronger predicates over \code{?g},
in the same way in which \code{spec[?h]} is being refined stepwise.
This is a possible final result:
\begin{alltt}
  (defun g (y1 y2) (append y1 y2))
  (defun-inst consp-io[g] (consp-io[?g] (?g . g)))
  (defthm member-of-append ; used to prove CONSP-IO[G]-LEMMA below
    (iff (member e (append y1 y2)) (or (member e y1) (member e y2))))
  (defthm consp-io[g]-lemma ; used to prove CONSP-IO[G]! below
    (implies (and (consp x) (io (car x) y1) (io (cdr x) y2))
             (io x (g y1 y2)))
    :hints (("Goal" :in-theory (disable io) :expand (io x (append y1 y2)))))
  (defthm consp-io[g]! (consp-io[g]) :hints (("Goal" :in-theory (disable g))))
\end{alltt}
The predicate \code{spec3[?h\us?f\us?g]} is strengthened
by replacing the \code{consp-io[?f]} conjunct
with one that constrains \code{?g} to be \code{g}:
\begin{alltt}
  (defun-sk2 def-?g (?g) ()
    (forall (y1 y2) (equal (?g y1 y2) (g y1 y2)))
    :rewrite :direct)
  (defun2 spec4[?h_?f_?g] (?h ?f ?g) ()
    (and (def-?h-fold[?f_?g]) (def-?f) (def-?g)))
  (defthm step4-lemma (implies (def-?g) (consp-io[?g]))
    :hints (("Goal" :in-theory '(consp-io[?g] consp-io[g]-necc
                                 consp-io[g]! def-?g-necc))))
  (defthm step4 (implies (spec4[?h_?f_?g]) (spec3[?h_?f_?g]))
    :hints (("Goal" :in-theory '(spec3[?h_?f_?g] spec4[?h_?f_?g] step4-lemma))))
\end{alltt}

\paragraph{Step 5}

Substituting the solutions \code{f} and \code{g} into \code{fold[?f\us?g]}
yields a solution for \code{?h}:
\begin{alltt}
  (defun-inst h (fold[?f_?g] (?f . f) (?g . g)))
  (defun-sk2 def-?h (?h) () (forall x (equal (?h x) (h x))) :rewrite :direct)
\end{alltt}
The conjunct \code{def-?h-fold[?f\us?g]} of \code{spec4[?h\us?f\us?g]}
is replaced with \code{def-?h},
which is equivalent to \code{def-?h-fold[?f\us?g]}
given the conjuncts \code{def-?f} and \code{def-?g}:
\begin{alltt}
  (defun2 spec5[?h_?f_?g] (?h ?f ?g) () (and (def-?h) (def-?f) (def-?g)))
  (defthm step5-lemma
    (implies (and (def-?f) (def-?g)) (equal (h x) (fold[?f_?g] x)))
    :hints (("Goal" :in-theory '(h fold[?f_?g] def-?f-necc def-?g-necc))))
  (defthm step5 (implies (spec5[?h_?f_?g]) (spec4[?h_?f_?g]))
    :hints (("Goal" :in-theory '(spec4[?h_?f_?g] spec5[?h_?f_?g]
                                 def-?h-fold[?f_?g] def-?h-necc step5-lemma))))
\end{alltt}
This concludes the derivation:
\code{spec[?h\us?f\us?g]} provides executable solutions
for \code{?h}, \code{?f}, and \code{?g}.
The resulting implementation is $\langle\code{h},\code{f},\code{g}\rangle$.
Chaining the implications of the step correctness theorems
shows that these solutions satisfy the requirements specification:
\begin{alltt}
  (defthm chain[?h_?f_?g] (implies (spec5[?h_?f_?g]) (spec[?h]))
    :hints (("Goal" :in-theory '(step1 step2 step3 step4 step5))))
\end{alltt}
More explicitly, instantiating the end-to-end implication shows that \code{h}
satisfies the requirements specification:
\begin{alltt}
  (defun-inst def-h (def-?h (?h . h)))
  (defun-inst def-f (def-?f (?f . f)))
  (defun-inst def-g (def-?g (?g . g)))
  (defun-inst spec5[h_f_g] (spec5[?h_?f_?g] (?h . h) (?f . f) (?g . g)))
  (defun-inst spec[h] (spec[?h] (?h . h)))
  (defthm-inst chain[h_f_g] (chain[?h_?f_?g] (?h . h) (?f . f) (?g . g)))
  (defthm spec5[h_f_g]! (spec5[h_f_g])
    :hints (("Goal" :in-theory '(spec5[h_f_g]))))
  (defthm spec[h]! (spec[h])
    :hints (("Goal" :in-theory '(chain[h_f_g] spec5[h_f_g]!))))
\end{alltt}

\section{Related Work}

The \code{instance-of-defspec} tool~\cite{joosten-reusing}
and the \code{make-generic-theory} tool~\cite{mateos-multiset}
automatically generate instances of functions and theorems
that reference functions constrained via encapsulation~%
\cite{kaufmann-structured},
by replacing the constrained functions
with functions that satisfy the constraints.
The instantiation mechanisms of these tools are similar to the ones of \soft;
constrained functions in these tools parallel function variables in \soft.
However, in \soft\
function variables are unconstrained;
constraints on them are expressed
via second-order predicates (typically with quantifiers),
and the same function variables can be used as parameters
of different constraining predicates.
Unlike \soft,
\code{instance-of-defspec} and \code{make-generic-theory}
do not handle choice and quantifier functions,
and do not generate termination proofs for recursive function instances.
\soft\ generates one function or theorem instance at a time,
while \code{instance-of-defspec} and \code{make-generic-theory}
can generate many.
These two tools are more suited to
developing and instantiating abstract and parameterized theories;
\soft\ is more suited to mimic second-order logic notation.

The \code{:consider} hint~\cite{moore-instantiation}
heuristically generates functional instantiations
to help prove given theorems.
\soft\ generates function and theorem instances
for given replacements of function variables;
from these replacements,
the necessary functional instantiations are generated automatically.

The \code{def-functional-instance} tool in the ACL2 community books
generates theorem instances for given replacements of functions.
This tool has more general use than \soft's \code{defthm-inst},
but it requires a complete functional instantiation,
while \code{defthm-inst} only requires replacements for function variables.

Wrapping existing events to record information for later use
(as done by \soft's
\code{defunvar}, \code{defun2}, \code{defchoose2}, and \code{defun-sk2})
has precedents.
For example, the \code{def:un-sk} tool~\cite{greve-quantified}
is a wrapper of \code{defun-sk} that records information
to help prove theorems involving quantifiers.
It may be useful to combine \code{def:un-sk}
with \soft's \code{defun-sk2} wrapper.

There are several tools
to generate functions and theorems according to certain patterns,
such as \code{std::deflist} in the ACL2 standard library
and \code{fty::deflist} in the FTY library~\cite{swords-fty}.
These tools may use \soft\
to generate some of the functions and theorems
as instances of pre-defined second-order functions and theorems.

A general-purpose theorem prover like ACL2
can represent a variety of specification and refinement formalisms,
e.g.~\cite{abadi-refinement,%
abrial-b,%
hoare-data,%
jones-vdm,%
milner-simulation,%
morgan-refinement,%
spivey-z};
derivations can be carried out within the logic.
But given the close ties to Applicative Common Lisp,
a natural approach to program refinement in ACL2
is to specify requirements on one or more target ACL2 functions,
and progressively strengthen the requirements
until the functions are executable and performant.

Alternatives to \soft's second-order predicates,
for specifying requirements on ACL2 functions,
include
\code{encapsulate}
(possibly via the wrappers \code{defspec} and \code{defabstraction}
in the ACL2 community books),
\code{defaxiom},
and \code{defchoose}.
But these are not as suited to program refinement:
\begin{itemize}
\item
  An \code{encapsulate} involves
  exhibiting witnesses to the consistency of the requirements,
  which amounts to writing an implementation and proving it correct.
  But it is the purpose of program refinement
  to construct an implementation and its correctness proof.
\item
  A \code{defaxiom} obviates witnesses but may introduce logical inconsistency.
\item
  A \code{defchoose} obviates witnesses and is logically conservative, but:
  \begin{itemize}
  \item
    It expresses requirements on single functions,
    necessitating the combination of multiple target functions into one.
  \item
    It expresses requirements
    on function results (the bound variables)
    with respect to function arguments (the free variables),
    but not requirements involving different results and different arguments,
    such as injectivity,
    non-interference~\cite{goguen-noninterference},
    and other hyperproperties~\cite{clarkson-hyperproperties}.
  \item
    It prescribes underspecified but fixed function results.
    For example, there is no clear refinement relation between
    the function introduced as
    \codelist{\code{defchoose}
              \code{f}
              \codelist{\code{y}}
              \codelist{\code{x}}
              \codelist{\code{>} \code{y} \code{x}}}
    and the function introduced as
    \codelist{\code{defun}
              \code{g}
              \codelist{\code{x}}
              \codelist{\code{+} \code{x} \code{1}}}.
  \end{itemize}
\end{itemize}
In contrast,
a second-order predicate can specify
any kind of requirements,
on multiple functions,
maintaining logical consistency,
and doing so without premature witnesses.

In the derivation in \secref{sec:progref-ref},
the use and instantiation of the generic folding function on binary trees
is an example of the application of algorithm schemas in program refinement,
as in~\cite{smith-mechanizing}
but here realized via second-order functions and theorems.
Second-order functions express algorithm schemas,
and second-order theorems show their correctness
under suitable conditions on the function parameters.
Applying a schema
adds a constraint that defines a target function to use the schema,
and introduces simpler target functions
corresponding to the function parameters,
constrained to satisfy the conditions for the correctness of the schema.

A refinement step from a specification \specN{j} can be performed manually,
by writing down \specN{j+1}
and proving
\codelist{\code{implies} \codelist{\specN{j+1}} \codelist{\specN{j}}}.
It is sometimes possible to generate \specN{j+1} from \specN{j},
along with a proof of
\codelist{\code{implies} \codelist{\specN{j+1}} \codelist{\specN{j}}},
using automated transformation techniques.
Automated transformations may require
parameters to be provided and applicability conditions to be proved,
but should generally save effort
and make derivations more robust against changes in requirements specifications.
At Kestrel Institute, we are developing
ACL2 libraries of automated transformations for program refinement.

\section{Future Work}

\paragraph{Guards}

\code{defun-inst} could be extended with the option
to override the default guard \instof{\icode{guard}}
with a different \icode{guard'},
generating the proof obligation
\codelist{\code{implies} \icode{guard'} \instof{\icode{guard}}}.
This would be useful in at least two situations.

A first situation is when
the function instance has more guard conditions to verify
than the second-order function being instantiated,
due to the replacement of a function parameter (which has no guards)
with a function that has guards.
Providing a stronger guard to the function instance
would enable the verification of the additional guard conditions.
For example,
an instance \code{quad[cdr]} of \code{quad[?f]} from \secref{sec:plain}
could be supplied with the guard \codelist{\code{true-listp} \code{x}}.

A second situation is when
the guard of the second-order function being instantiated
includes conditions on function parameters that involve a quantifier,
e.g.\ the condition that the binary operation \code{?op}
of a generic folding function over lists
is closed over the type \code{?p} of the list elements.
Instantiating \code{?p} with \code{natp} and \code{?op} with \code{binary-+}
satisfies the condition,
but \instof{\icode{guard}} still includes a quantifier
that makes the instance of the folding function non-executable.
Supplying a \icode{guard'} that rephrases \instof{\icode{guard}}
to omit the satisfied closure condition would solve the problem.
As guard obligations on individual parameters are relieved
when functions are applied to terms in a term,
it makes sense to relieve guard obligations on function parameters
when second-order functions are ``applied'' to functions in \code{defun-inst}.

\code{defun-inst} could also be extended with the ability
to use the instances of the verified guard conditions
of the second-order function being instantiated,
to help verify the guard conditions of the function instance.
This may completely verify the guards of the instance,
when no guard overriding is needed.

\paragraph{Partial Functions}

\soft\ could be extended with a macro \code{defpun2}
to introduce partial second-order functions,
mimicked by partial first-order functions
introduced via \code{defpun}~\cite{manolios-partial}.
\code{defun-inst} could be extended to generate
not only partial function instances,
but also total function instances
when the instantiated \code{:domain} or \code{:gdomain} restrictions
are theorems.
Partial second-order functions would be useful, in particular,
to define recursive algorithm schemas
whose measures and whose argument updates in recursive calls
are, or depend on, function parameters.
An example is a general divide-and-conquer schema.%
\footnote{The folding function from \secref{sec:plain}
is a divide-and-conquer schema specialized to binary trees.}

\paragraph{Mutual Recursion}

\soft\ could be extended with a macro \code{mutual-recursion2}
to introduce mutually recursive plain second-order functions
(with \code{defun2}),
mimicked by mutually recursive first-order functions
introduced via \code{mutual-recursion}.
\code{defun-inst} could be extended
to generate instances of mutually recursive second-order functions.

\paragraph{Lambda Expressions}

\code{defun-inst} and \code{defthm-inst} could be extended
to accept instantiations that map function variables to lambda expressions,
similarly to \code{:functional-instance}.

\paragraph{Instantiation Transitivity}

If \sofunP\ is introduced as the \inst\ instance of \sofun,
and \fun\ is introduced as the \inst' instance of \sofunP,
then \fun\ should be the \inst'' instance of \sofun,
where \inst'' is a suitably defined composition of \inst\ and \inst'.
Currently \code{defun-inst} does not record \fun\ as an instance of \sofun\
when \fun\ is introduced,
but it could be extended to do so.
With this extension, \code{injective[quad[wrap]]} in \secref{sec:sothminst}
would be the \codelist{\codecons{\code{?f}}{\code{quad[wrap]}}} instance
of \code{injective[?f]} in \secref{sec:quant}.

In a related but different situation,
given \sofun, \sofunP, \fun, \inst, \inst', and \inst'' as above,
but with \fun\ introduced as the \inst'' instance of \sofun,
and \sofunP\ introduced as the \inst\ instance of \sofun,
in either order (i.e.\ \fun\ then \sofunP, or \sofunP\ then \fun),
then \fun\ should be the \inst' instance of \sofunP.
Currently \code{defun-inst} does not record \fun\ as an instance of \sofunP\
when \fun\ (after \sofunP) or \sofunP\ (after \fun) is introduced,
but could be extended to do so.
With this extension,
if \code{injective[quad[wrap]]} were introduced
as the \codelist{\codecons{\code{?f}}{\code{quad[wrap]}}} instance
of \code{injective[?f]},
and \code{injective[quad[?f]]} were introduced
as the \codelist{\codecons{\code{?f}}{\code{quad[?f]}}} instance
of \code{injective[?f]} as in \secref{sec:sofuninst},
then \code{injective[quad[wrap]]} would be
the \codelist{\codecons{\code{?f}}{\code{wrap}}} instance
of \code{injective[quad[?f]]}.

An alternative to these two extensions of \code{defun-inst}
is to extend \soft\ with a macro to claim that
an existing instance of a second-order function
is also an instance of another second-order function
according to a given instantiation.
The macro would check the claim
(by applying the instantiation and comparing the result with the function)
and extend the table of instances of second-order functions
if the check succeeds.
In the first scenario above,
the macro would be used to claim that \fun\ is the \inst'' instance of \sofun;
in the second scenario above,
the macro would be used to claim that \fun\ is the \inst' instance of \sofunP.

\paragraph{Function Variable Constraints}

Currently the only constraints on function variables are their types.
\code{defunvar} could be extended
to accept richer signatures for function variables,
with multiple-value results and single-threaded arguments and results.
\code{defun-inst} and \code{defthm-inst} would then be extended
to check that instantiations satisfy these additional constraints.
A more radical extension would be
to attach logical constraints to certain function variables,
as in encapsulations.

\paragraph{Automatic Instances}

As explained in \secref{sec:sofuninst},
when an instantiation is applied to a term,
the table of instances of second-order functions is consulted
to find replacements for certain second-order functions,
and the application of the instantiation fails if replacements are not found.
Thus, all the needed instances must be introduced
before applying the instantiation,
e.g.\ in \secref{sec:sothminst}
the two \code{defun-inst}s had to be supplied
before the last \code{defthm-inst}.
\soft\ could be extended
to generate automatically the needed instances of second-order functions.

\soft\ could also be extended with a macro \code{defthm2}
to prove a second-order theorem via \code{defthm}
and to record the theorem in a table,
along with information about the involved second-order functions.
\code{defun-inst} could be extended with the option to generate
instances of the second-order theorems
that involve the second-order function being instantiated.
\code{defthm2} could include the option
to generate instances of the theorem
that correspond to the known instances of the second-order functions
that the theorem involves.
These extensions would reduce the use of explicit \code{defthm-inst}s.

The convention of including function variables in square brackets
in the names of second-order functions and theorems,
could be exploited to name
the automatically generated function and theorem instances,
as suggested by the examples throughout the paper.

\paragraph{Other Events}

\soft\ could be extended to provide second-order counterparts
of other function and theorem introduction events,
e.g.\ \code{define}, \code{defines}, and \code{defrule}
in the ACL2 community books.

\bibliographystyle{eptcs}

\bibliography{paper}

\end{document}